\documentclass[showpacs,preprintnumbers,amsmath,amssymb]{revtex4}

\usepackage{graphicx}
\usepackage{dcolumn}
\usepackage{bm}
\begin{document}

\preprint{FNT/T-2004/21}

\title{Elementary particles, holography and the BMS group}

\author{Claudio Dappiaggi}
\email{claudio.dappiaggi@pv.infn.it}
\affiliation{Dipartimento di Fisica Nucleare e Teorica\\
Universit\`{a} degli Studi di Pavia, INFN, Sezione di Pavia, \\
via A. Bassi 6, I-27100 Pavia, Italy
}


\date{\today}

\begin{abstract}
In the context of asymptotically flat space-times, it has been suggested to label
elementary particles as unitary irreducible representations of the BMS group. We
analyse this idea in the spirit of the holographic principle advocating the
use of this definition.
\end{abstract}

\pacs{04.62.+v, 04.20.Ha, 11.30.Cp, 04.60-m}
\maketitle
The concept of elementary particle plays a central role in the physical
interpretation of quantum field theory; nonetheless we still lack a concrete and
universally accepted definition whenever gravity is included and, thus, a 
non trivial background space-time is considered. The aim of this letter is to advocate
that, in the framework of four-dimensional asymptotically flat space-times, a solution to this
deficiency exists if the overall problem is set in the context of finding an holographic
description for a quantum field theory in such class of space-times, \cite{'thooft}.

As a starting point, let us remember that, thought as \cite{Weinberg,Barut}, a system is
``elementary''  when its Hilbert space carries a single irreducible representation (irrep.) of (the 
double cover of) the full Poincar\'e group $P=SL(2,\mathbb{C})\ltimes T^4$, the semidirect product between the four dimensional translations $T^4$ and $SL(2,\mathbb{C})$, the group of conformal motions of the 2-sphere $S^2$.
Since an elementary system may admit under internal probing a more complex
structure, an \emph{elementary particle} is defined as an elementary system
whose states cannot be physically connected to states of another system.
Albeit natural,
the above definition is unsatisfactory for several reasons \cite{Mc10, Barut}: 
1) in a Poincar\'e invariant theory, the mass operator admits only a continuous spectrum
whereas observations show only a discrete spectrum of (rest) masses which
cannot be described by any finite dimensional Lie group \cite{Raifeartaigh},
2) by means of the Wigner approach, it is possible to construct all
the kinematical and the dynamical data of a Poincar\'e free field theory, but 
massless particles may be labelled either by discrete spins with a finite
number of polarizations (unfaithful representations) either by continuous spins
with an infinite number of polarizations (faithful representations); only the
former have been experimentally observed though there is no theoretical reason
to prefer any of the above two choices,
3) more importantly, the definition of elementary particles assumes the flatness of the
background discarding any gravitational effect. In a general relativity
framework, even in presence of a weak gravitational field, this is not a
reasonable request: Poincar\'e invariance is assumed on
the basis that the underlying manifold is maximally symmetric i.e., it is 
Minkowski whereas, according to Einstein's theory, the degree of symmetry of any 
other bulk space-time is smaller. 

A candidate solution for the above pathologies can be formulated in the context 
of asymptotically flat (AF) space-times where a natural and universal
counterpart for the $P$ group exists \cite{Mc10}. In detail, all AF manifolds share a common boundary structure at past and 
future null infinity. In a Bondi reference frame $(u=t-r,r,\theta,\varphi)$, 
these submanifolds, $\Im^\pm$, topologically equivalent to
$S^2\times\mathbb{R}$, can be endowed with a degenerate metric 

$$ds^2=0\cdot du^2+d\theta^2+\sin^2\theta d\varphi^2,$$
whose group of 
diffeomorphisms is the so-called Bondi-Metzner-Sachs group (BMS) which, up to a
stereographic projection sending $(\theta,\varphi)$ in $(z,\bar{z})$, is 
\begin{gather}
u\to u^\prime=K(z,\bar{z})[u+\alpha(z,\bar{z})],\\
z\to z^\prime=\frac{az+b}{cz+d}\;\;\;ad-bc=1\wedge a,b,c,d\in\mathbb{C},
\end{gather} 
where $K(z,\bar{z})=(1+\mid z\mid^2)^{-1}(\mid az+c\mid^2+\mid bz+d\mid^2)$ and
where $\alpha(z,\bar{z})$ is an arbitrary real scalar function over $S^2$
\cite{Mc1}.
This transformation identifies the BMS as 
the semidirect product $SL(2,\mathbb{C})\ltimes N$ where $N$ is the set of
$\alpha$-functions endowed with a suitable topology usually, but not 
necessary, chosen as $N=L^2(S^2)$ i.e. the collection of square integrable maps
over the 2-sphere \cite{Girardello, Crampin}.
The universality of the boundary structure and the dual role of the BMS group
as diffeomorphism group of $\Im^\pm$ and as asymptotic symmetry group
of any AF bulk metric naturally suggests to replace the Poincar\'e group with the
BMS as the fundamental group of symmetry; thus, an elementary particle in
an AF space-time is defined by means of an elementary system whose Hilbert space carries
a unitary irreducible representation of $SL(2,\mathbb{C})\ltimes L^2(S^2)$.
Although this approach experienced an initial success, no
further significant progress was achieved in this field after McCarthy analysis of BMS
theory of induced representations. In \cite{Mc1}, it was pointed out that,
besides unitary irrep. related to the observed Poincar\'e massive and massless
fields, a plethora of other elementary particles existed, so far lacking any
experimental evidence. As we have anticipated, considering McCarthy seminal work
as a starting point, we
will nonetheless advocate the effectiveness of the whole approach; in particular
the unwanted pathologies dissapear if we interpret the BMS field theory as a
boundary theory encoding holographically the information from \emph{any
asymptotically flat} space-time.

Let us briefly comment that holography has been introduced
in order to solve the apparent paradox information of black holes by means of
a second theory living in a codimension one submanifold (usually the boundary)
with a density of data not exceeding the Planck density. An explicit realization of
these concepts consists on constructing a field theory on the boundary of
the chosen space-time invariant under the action of the asymptotic symmetry
group; the bulk data are reconstructed starting directly from those associated
to the
boundary, explaining how they generate their dynamic and how they can reproduce
classical space-time. A concrete example is known in an AdS manifold 
as the AdS/CFT correspondence \cite{ADS/CFT} and, only recently, a similar
investigation has begun in the context of AF space-times \cite{Arcioni}. In this
latter scenario, the aim is to develop a field theory on $\Im^\pm$ invariant
under a BMS transformation and, consequently, McCarthy analysis can be naturally
interpreted as the initial framework where the boundary kinematic data are
studied and classified. In particular, adopting notations and nomenclatures as
in \cite{Mc1}, a \emph{BMS covariant wave function(al)} is defined as a map
\begin{equation}\label{covbms}
\psi: L^2(S^2)\longrightarrow\mathbb{C}^n,
\end{equation} 
transforming under a BMS unitary representation, i.e., in a momentum frame and for
any $g=(\Lambda,p(\theta,\varphi))\in BMS$,
\begin{equation}\label{waveq}
\left(D^\lambda(g)\psi\right)(p^\prime)=e^{i<p,\alpha>_{L^2(S^2)}}U_\lambda(\Lambda)\psi(\Lambda^{-1}p^\prime),
\end{equation}
being $U_\lambda(\Lambda)$ a unitary representation of $SL(2,\mathbb{C})$ and
$e^{i<p,\alpha>_{L^2(S^2)}}$ the character associated to $p(\theta,\varphi)$.
In order to describe an elementary particle (or equivalently a free
field), the associated wave function should transform under a unitary and
\emph{irreducible} representation; the latter can be constructed
from the unitary representation of the little (isotropy) groups $L\subset BMS$
and consequently it is possible to introduce an \emph{induced wave function}:
\begin{equation}
\tilde{\psi}:\mathcal{O}_L\sim\frac{SL(2,\mathbb{C})}{L}\longrightarrow\mathbb{C}^m,\;\;m<n
\end{equation}
transforming under an irrep. of BMS
induced from one of $L$. Thus, according to this setting, the kinematic data for
each elementary particle are fully characterized by $L$ and by the associated
Casimir invariant, the squared mass $m^2$ of the free fields. Let us stress
that the set of possible little groups includes $SU(2)$ with $m^2>0$,
$\Gamma$, the double cover of $SO(2)$, with $m^2=0$ and a plethora of other non-connected
isotropy subgroups, the most notables being the series of finite alternate,
cyclic and dihedral groups $A_n$,
$C_n$, $D_n$ with $n>2$. The connected little groups provide exactly the unitary
irrep. giving rise to the observed Poincar\'e spins; as a direct consequence, the
arbitrariness in the choice of the irrep. associated to massless particles
disappears since the faithful one-dimensional representation proper of the BMS 
$\Gamma$ little group is fully equivalent to its Poincar\'e counterpart induced 
from the two dimensional euclidean group $E(2)\subset P$.\\
The main handicap emerging from the analysis of the kinematic data is
the total absence of an interpretation for the additional ``non-Poincar\'e''
degrees of freedom. The paradigm we propose is the
following: if a BMS field theory
encodes the data from \emph{all} AF manifolds, an
elementary particle, living in a fixed background, such as, for example, 
Minkowski space-time, is described only by means of those boundary degrees of freedom 
allowing a proper reconstruction of the chosen bulk manifold.\\
In order to support such conjecture, the first step is to compare
the dynamic of bulk and boundary free fields. In the context of
Wigner approach, the latter can
be fully characterized as a set of constraints restricting the covariant wave
function to the induced one; in particular, in a BMS setting, starting from
(\ref{covbms}), these constraints are twofolds: the first is
\begin{equation}\label{orthop}
\rho(p)\psi(p)=\psi(p),
\end{equation}
an orthoprojection equation where $\rho(p)$ is a suitable non a priori
invertible covariant operator which cancels the redundant component of $\psi(p)$
in $\mathbb{C}^n$ i.e. the image of $f$ is (isomorphic to) $\mathbb{C}^m$. The second is an
orbit constraint that reduces the support of (\ref{covbms}) from $L^2(S^2)$ to
the coset space $\mathcal{O}_L$; although an explicit expression is available
for all little groups \cite{Arcioni}, we switch for sake of clarity to a 
specific example: $L=SU(2)$ where the orbit equation is
\begin{equation}\label{BMSKG}
[\eta^{\mu\nu}\pi(p)_\mu\pi(p)_\nu-m^2]\psi(p)=0,\;\;\;[p-\pi(p)]\psi(p)=0,
\end{equation}
being $\pi(p)$ the so-called Poincar\'e momentum i.e. a vector constructed by the
first four coefficients in the spherical harmonic expansion of each
$p(\theta,\varphi)$ 
$$\pi(p)_\mu=\pi\left(\sum\limits_{l=0}^\infty\sum\limits_{l=-m}^m
p_{lm}Y_{lm}(\theta,\varphi)\right)=(p_{00},...,p_{11}).$$
Let us emphasize that, while (\ref{BMSKG}) is the BMS-equivalent of the
Klein-Gordon equation which holds for any Poincar\'e covariant elementary 
particle, (\ref{orthop}) is a compact expression for the wave equations
of any BMS free field i.e. they are the BMS
counterpart for usual formulas such as as the Dirac and the Proca wave 
equations. Thus, following this line of reasoning, the pair
$\left\{D^\lambda(\Lambda),\rho(p)\right\}$ (from (\ref{waveq}) and
(\ref{orthop})) completely characterise the dynamic
of a free field; each (BMS) elementary particle is distinguished from another
only by the values of the squared mass and of the spin.\\
Since, according to the holographic principle, the boundary theory should encode the
bulk degrees of freedom, a comparison, between the classical dynamic of 
a theory living on $\Im^\pm$ and of one living on a flat background, should be
performed at a level of phase spaces. The subtlety lies in the intrinsic
infinite dimensional nature of the BMS field theory which prevents a canonical
approach to the construction of the phase space since the usual splitting of a four
dimensional manifold $M^4$ as $\Sigma_3\times\mathbb{R}$ is meaningless in the
boundary framework. Thus we introduce the \emph{covariant phase space},
the set of covariant wave function(al)s satisfying the equations of motion and,
consequently, representing the dynamically allowed configurations;
in the specific example of a BMS $SU(2)$ field it is 
\begin{equation}\label{BMScov}
\Gamma^{(cov)}_{BMS}=\left\{\psi:L^2(S^2)\to\mathbb{R},[\pi(p)^\nu\pi(p)_\nu-m^2]\psi(p)=0,
[p-\pi(p)]\psi(p)=0,\;\;\;\rho(p)\psi(p)=\psi(p)\right\}.
\end{equation}
The Poincar\'e counterpart of this expression for an $SU(2)$ field is:
\begin{equation}\label{Pcov}
\Gamma^{(cov)}_{P}=\left\{\psi:T^4\to\mathbb{R}, [p^\mu p_\mu-m^2]\psi(p)=0,\;\;\rho(p^\mu)\psi(p^\mu)=\psi(p^\mu)\right\}.
\end{equation} 
It is straightforward to realize that, due to the orbit constraint
$[p-\pi(p)]\psi(p)=0$, (\ref{BMScov}) is in 1:1 correspondence with (\ref{Pcov});
furthermore, an identical claim holds 
between the covariant phase space of a Poincar\'e $E(2)$
massless field and a BMS $\Gamma$ massless particle with vanishing pure
supertranslational component \cite{Arcioni2}. 
In an ``holographic''
language, this result grants us that the
boundary theory fully encodes the bulk classical degrees of freedom (at least in
Minkowski); conversely, from an ``elementary particle'' point of view, the 
results from \cite{Mc10} are considerably improved since, not only the
kinematic but also the \emph{dynamic} of massive and massless Poincar\'e
elementary particles is fully reproduced in a BMS invariant theory.

As a final step, we need to provide evidences that all other BMS little group do
not encode any information allowing a full reconstruction of the physics and the
geometry of a Minkowski space-time. A solution to this
obstacle lies in the so called null
surface formulation of general relativity. In this approach to Einstein's
theory, the main variable is a scalar function $Z:M^4\times S^2\to\mathbb{R}$
(cut function) solution of the light cone equation in $M^4$ \cite{Frittelli};
$Z(x_a,\theta,\varphi)$ allows to univocally reconstruct all the conformal data of the bulk
manifold and, in particular, up to a conformal rescaling the metric itself. From
an holographic perspective, the appealing aspect of the overall procedure arises
realizing that, helding fixed the bulk point $x_a\in M^4$, the cut function is a
real scalar map on $\Im^\pm$, thus a boundary data. Moreover since
$Z_{x_a}$ is smooth and single-valued in a suitable neighbourhood of $\Im^\pm$,
it can be naturally identified as a BMS supertranslation. Thus, in a BMS field theory, the collection of data encoding the free fields 
dynamic on a fixed background, can be 
extracted from the degrees of freedom $\left\{Z_{x_a}\right\}$ reconstructing a 
specific manifold in the null surface formalism. 

Within this framework, the
set of cut functions appears to play a role similar to the Fefferman-Graham construction
for an asymptotically AdS space-time (see for example \cite{Skenderis, deHaro});
this latter tool allows for an \emph{algebraic} reconstruction of bulk data
starting from boundary ones whereas the counterpart of this approach in an
asymptotically flat space-time produces a set of \emph{differential} equations
\cite{deHaro2}. Conversely, the null surface formulation of general relativity
and, more in detail, $Z_{x_a}(z,\bar{z})$ allows in an asymptotically flat
space-time a reconstruction of bulk geometry (in particular the metric) starting only 
from data living on $\Im^+$ (or $\Im^-$) solving a set of algebraic equations.
Furthermore also bulk fields on $M^4$ can be seen as ``dependant'' only upon boundary
data since, starting only from the cut functions, it is possible to construct the
following tetrad $\Theta^i$ living at null infinity:
$$ u=Z(x_a,z,\bar{z}),\;\omega=(1+|z|^2)\partial
Z(x_a,z,\bar{z}),\;\bar{\omega}=(1+|z|^2)\bar{\partial}Z(x_a,z,\bar{z}),\;R=(1+|z|^2)^2
\partial\bar{\partial}Z(x_a,z,\bar{z}),$$ 
which can be (in principle) inverted as $x_a=x_a[\Theta^i,z,\bar{z}]$. Thus
each local bulk field $\phi^\lambda:M^4\to\mathbb{C}^\lambda$ can be now rewritten
as a functional of boundary data i.e. $\phi^\lambda(x_a)=\phi^\lambda[\Theta^i,z,\bar{z}]$.
  
In particular, if we now consider the specific example of a Minkowski background and 
if we work in a momentum frame, the cut function is unique \cite{Kozameh}:
\begin{equation}\label{cut}
Z_{p_a}(\theta,\varphi)=p(\theta,\varphi)=p_al^a(\theta,\varphi),
\end{equation}
where $l^a=\left\{Y_{00}(\theta,\varphi),...,Y_{11}(\theta,\varphi)\right\}$.
At a classical level, (\ref{cut}) grants us that the momenta encoding the
information from a flat manifold automatically satisfy a vanishing pure
supertranslational constraint
\begin{equation}\label{cut2}
Z_{p_a}-\pi(Z_{p_a})=0,\;\textrm{i.e.}\;\;p-\pi(p)=0
\end{equation}
Thus a BMS elementary particle can be related
to a Poincar\'e invariant counterpart living in Minkowski only if the equation
of motion for the associated covariant wave functional (\ref{BMScov}) includes 
(\ref{cut2}). For a fixed little group $L$, the orbit equation 
imposes to the classical free field an evolution on a finite dimensional
manifold embedded in $L^2(S^2)$; the latter is
generated by the action of the coset group $\frac{SL(2,\mathbb{C})}{L}$ on a
fixed point $\bar{p}\in L^2(S^2)$ such that $L\bar{p}=\bar{p}$. A decomposition in spherical
harmonics proves that the most general expression for $\bar{p}$ is \cite{Mc1, Arcioni}
\begin{gather}\label{prima}
\bar{p}=m+\sum\limits_{l>1}\sum\limits_{m=-l}^lp_{lm}Y_{lm}(\theta,\varphi),\\
\bar{p}=p_0+p_0Y_{11}(\theta,\varphi)+\sum\limits_{l>1}\sum\limits_{m=-l}^lp_{lm}Y_{lm}(\theta,\varphi),\label{seconda}
\end{gather}
respectively for a massive and a massless field. From the above two formulas, it
is straightforward to see that
(\ref{cut2}) is equivalent to the constraint $p_{lm}=0$ for any $l>1$. 
A detailed analysis, \cite{Arcioni, Arcioni2}, proved that this 
request may be satisfied only by the connected isotropy subgroups of the BMS 
group i.e. $SU(2)$, if we consider (\ref{prima}), and $\Gamma$ if we consider
(\ref{seconda}); thus, we may conclude that, at least at a classical level, only 
these two BMS subgroups encode the information from a Minkowski
elementary particle, discarding any physical role for the other ``pathological'' little
groups. 

In order to better clarify the role of the BMS group in the definition of
elementary particles, we need to comment on the
quantum aspects of the boundary theory. In the above framework, we
focused our attention mainly on the dynamically allowed configurations whereas, 
if we wish to calculate quantum data, such 
as correlation functions, by means of path-integral techniques, we should refer
to all the kinematically allowed configurations. The latter are a priori different
in a Poincar\'e and in a BMS field theory and it is natural to wonder if the
conjectured correspondence holds also at this level. Thus we
need to switch to a Lagrangian formalism; if we consider for sake of
simplicity a BMS scalar field $\phi(x)$, the equation of motion (\ref{BMSKG}) can be
derived minimizing the following action \cite{Dappiaggi}
\begin{equation}
S(\phi)=\int\limits_{L^2(S^2)}d\mu\left\{\phi(x)[\eta^{\mu\nu}D_{e_\mu}D_{e_\nu}-m^2]\phi(x)+\sum\limits_{i=1}^\infty\gamma_i(x)D_{e_i}\phi(x)\right\},
\end{equation}
being $e_\mu$ an element of the set
$\left\{Y_{00}(\theta,\varphi),...,Y_{11}(\theta,\varphi)\right\}$, $e_i$ one of
the set $\left\{Y_{lm}(\theta,\varphi)\right\}_{l>1}$, $D_{e_i}$ the infinite
dimensional directional derivative along $e_i$ and $\gamma_i(x)$ a Lagrange
multiplier. The corresponding partition function is 
\begin{gather}
S=\int\limits_{\mathcal{C}}d[\phi]e^{iS(\phi)}=const\cdot
det[B]^{-\frac{1}{2}},\\
B=\eta^{\mu\nu}D_{e_\mu}D_{e_\nu}+m^2+\sum\limits_{i=1}^\infty\frac{1}{2\zeta_i}(Q_{e_i}-D_{e_i})D_{e_i},
\end{gather}
where $\zeta_i$ is an arbitrary real non vanishing number and where $Q_{e_i}$ is
the infinite dimensional multiplication operator along the direction $e_i$. The
propagator $G(x_1-x_2)$ can be calculated as 
\begin{equation}\label{prop}
BG(x_1-x_2)=i\delta(x_1-x_2).
\end{equation}
Up to a Fourier transform, (\ref{prop}) satisfies:
\begin{equation}\label{propagator}
[\eta^{\mu\nu} p_\mu
p_\nu-m^2+\sum\limits_{i=1}^\infty(p_iD_{e_i}-p^2_i)]G\left(p(\theta,\varphi)\right)=i,
\end{equation}
where $p_\mu$ and $p_i$ are the projections of $p(\theta,\varphi)$ respectively
along the directions $e_\mu$ and $e_i\in L^2(S^2)$.
A physical analysis of (\ref{propagator}) has been performed in
\cite{Dappiaggi}, but, in this letter, we wish simply to emphasize the relation
of the above formula with the flat counterpart i.e., if we 
take into account (\ref{cut}) as the set of possible values of
$p(\theta,\varphi)$, (\ref{propagator}) reduces to 
$$G(p)=\frac{i}{\eta^{\mu\nu} p_\mu p_\nu-m^2},$$ 
which is the 2-point function in a Minkowski background. 
Thus, this result suggests us that the conjecture to holographically describe 
Poincar\'e elementary particles by means of the BMS group should hold also at a 
quantum level.\\
To conclude this letter, we wish to emphasize some remarks on the overall
approach:\\
$\bullet$ in a general picture, elementary particles may also be
characterized by an additional set of quantum number $\left\{\sigma\right\}$ 
associated to internal degrees of freedom usually described by means of a
(gauge) Lie group $G$. Nonetheless the indices $\left\{\sigma\right\}$ act as 
absolute superselection rules i.e. external interactions can only modify the momentum
and the spin projection along a fixed direction. The suggestion in
\cite{Komar} to relate these degrees of freedom to the irrep. of $I=\frac{BMS}{T^4}$, does not seem to hold in an
holographic framework; on the contrary the faithful irrep. of $I$
label the so called IR-sectors of gravity \cite{Arcioni2}. In a few words, the
presence of different infrared sectors of the gravitational field is a measure
of the arbitrariness induced by the BMS group in the choice of a specific Minkowski 
space-time describing the underlying geometry of a bulk field approaching $\Im^\pm$. 
This specific degree of freedom is related to pure supertranslations and,
consequently, to the $I$ group; it represents a direct
consequence of the obstruction to reduce the BMS to the Poincar\'e group, 
thus it has no reference with internal labels of an elementary particle.\\
$\bullet$ the absence of a physical interpretation for the non connected little 
groups disappears in a generic (non stationary) background. Conversely, they may carry
information from specific bulk data and an example is provided by the discrete isotropy
subgroups, related to gravitational instantons (see \cite{Melas} and references
therein).\\ 
$\bullet$ a further question concerns the application of the hypothesis
proposed in this letter in a scenario with a non vanishing cosmological constant
and in particular in the AdS$_d$/CFT$_{d-1}$ ($d>3$) correspondence. Let us
briefly comment that, though completely different from its asymptotically flat
counterpart (formulated only in 4D), in an AdS manifold holography relies to a certain extent on the equivalence between the bulk and the boundary (finite
dimensional) symmetry group; thus, a priori, there is no specific reason that do
not allows to repeat the reasoning of this letter in such framework though a
detailed analysis is not yet available.\\
$\bullet$ the role of interactions both in bulk and in the boundary has been 
discarded in this letter
since our aim has been to develop an alternative definition of elementary particles
which are related to free fields. Nonetheless, in the spirit of finding an
holographic correspondence in asymptotically flat space-times, it is imperative
to understand the role of interactions between BMS fields and whether they may
``break the holographic machinery''. According to the initial analysis of the 
boundary theory in \cite{Dappiaggi}, the leading role played by cut functions in
the bulk reconstruction starting from boundary data, appears to hold even in
presence of interactions. A tricky issue arises if one wishes to consider
boundary gauge theory since the usual construction coupling gauge fields and
elementary particles, well explained in \cite{Weinstein}, cannot be blindly
applied in the infinite dimensional context proper of a BMS setting; thus this
issue is still under analysis and development. 
\vspace{-.31cm}

\begin{acknowledgments} 
 
The author is in debt with Mauro Carfora and Giovanni Arcioni for useful discussions and
comments.
\end{acknowledgments}

\end{document}